# “That’s AI Slop, You Bot!”: Studying Accusations, Evidence, and Credibility in Online Discourse Towards LLM-Generated Comments

Jason Miklian, University of Oslo
John E. Katsos, American University of Sharjah (corresponding author)

**Abstract**

Generative AI has made fluent prose cheap to produce, breaking the old promise to readers that good writing meant real thinking. How have readers responded, and what can this tell us about changing anti-AI attitudes? We analyzed 25 million comments from Hacker News and Reddit (2023-2026), combining LLM judgment on 7,500 sampled accusations of AI use, sentiment trajectories, speech-act coding of 300 confirmed accusations of AI use, and a matched-control test of accused versus non-accused parent comments. We found that the pejorative-label share of accusations rose more than tenfold on both platforms while a placebo vocabulary of pre-2022 inauthenticity terms (“shill”, “astroturf”) did not. This shift reflected a fast-growing trend of branding any suspicious or seemingly inauthentic prose as “AI slop”. The slop frame now constitutes 94 percent of pejorative mentions, with the dominant comments shifting in tone from mockery toward gatekeeping and structural protest. The key surprise comes from a matched-control test which found that prose features that statistically distinguish AI from human text do not predict which human text gets accused as AI. The new accusations work as social gatekeeping of perceived authenticity without actually screening for AI. This research extends signaling theory by showing that substitute signals used socially can grow even when inaccurate if the underlying detection problem cannot be solved at the non-expert level. It shows that AI's effects on writing from the reader side are distinct from those on the production (writer) side. Detection technology cannot resolve this dynamic because the social function of accusations is increasingly to perform social gatekeeping and in-group signaling as opposed to identifying AI-generated writing.

## 1. Introduction

The arrival of generative large language models (LLMs) has both cheapened the production and amplified the proliferation of fluent prose. Most work on AI and writing has tracked impacts on the producer side, asking whether writers changed their prose in response to AI tools, or if workers are hired less often because content no longer indicates worker quality reliably (Galdin & Silbert 2025). With the proliferation of Generative AI, the signal "good prose means real effort, thinking, and expertise" became cheap to fake at scale, with deep consequences for AI-influenced content markets (Howell & Potgieter 2025; Tullis 2025) and for the credibility of research that depends on written signals (Obschonka & Lévesque 2026).

Existing accounts of AI and writing largely treat the writer as the locus of action through avoidance, adoption, or hybridization. But how do consumers (readers) feel about this change? What can be learned by studying the actions of commenters who impose a cost on stigmatized features through accusation speech?

The article proceeds as follows. Section 2 develops the theoretical scaffolding. Employing signal theory, we ask whether and how consumers organize around harder to fake properties. We examine the possibility that AI use accusations that are unfounded might invert the direction of harm, where the model has been positioned as a perpetrator of testimonial injustice (Kay et al. 2024). Here, lay AI (il)literacy would be the agent, and the writer/producer is the target. If the social function of accusations of AI use is, in fact, gatekeeping rather than the identification of AI-generated writing, detection-improving technology may exacerbate the problem it intends to fix.

Section 3 describes our data and methods, and Section 4 reports the empirical findings that the trajectory has the shape that signaling theory predicts, namely, that when a trusted signal collapses, populations coordinate on a substitute signal that is harder to fake. The “slop” accusation has that shape: short, recognizable, costly to deploy sincerely, cheap for others to reproduce. We argue that debates about AI writing have focused too narrowly on production while overlooking how readers respond. Section 5 discusses the implications of the findings, suggesting that accusations like “AI slop” spread because they serve social functions such as boundary maintenance, status signaling, and gatekeeping rather than because they accurately detect AI-generated text. In other words, accusations are less a screening tool than a form of social regulation.

## 2. Theoretical framing

The standard prediction from signal theory would be that when exposed to LLM-heavy material, readers will coordinate on a more authentic substitute whose properties make it harder to fake than the original. Given that markets with information asymmetry between buyers and sellers tend toward degraded equilibria where bad-quality producers dominate (Akerlof 1970), and costly signals like credentials or warranties can re-coordinate buyers and sellers around a substitute that is harder to fake (Spence 1973), two questions follow from this framework.

The first is descriptive: *Do online readers coordinate on a substitute signal for human authorship after exposure to generative LLM content, and if so, what properties characterize it?* (RQ1). The second tests the framework's core prediction about what makes a substitute viable: *Does the substitute signal satisfy the screening-accuracy condition that signaling theory predicts is necessary for a counter-signal to survive?* (RQ2). If RQ1 confirms substitution but RQ2 returns a null, the case becomes theoretically productive because signal theory would then need to be supplemented from a different mechanism to explain why the substitute persists.

The mechanism by which a substitute stabilizes in a population in the context of online writing is sociolinguistic. Enregisterment theory describes how a set of features is recognized as a coherent variety indexing particular speakers or stances (Agha 2003, 2007), extended to cases ranging from Pittsburghese to Netspeak on timescales of decades or longer (Silverstein 2003;

Johnstone et al. 2006; Squires 2010). A complementary account describes how ordinary speakers participate in defining language through cumulative metalinguistic speech acts (Cameron 1995; Lukač & Heyd 2023). Stance theory describes the individual speech-act mechanism (Du Bois 2007; Bucholtz & Hall 2005). When a Reddit user replies to a post, "This is AI slop, get this off the sub," the speech act evaluates the writing, positions the speaker against the writer, and claims community membership and gatekeeping in one short move. Cumulative stance-taking by community members aggregates into what is called enregisterment.

The features of major platforms such as textual persistence, large audiences, ranking, and low cost of uptake accelerate the cycle from the multi-decade timescale of classical cases to whatever timescale the users engage on. The rapid adoption of explicit subreddit-level AI rules between mid-2023 and late 2024 (Lloyd et al. 2025) raises the possibility that accusation stabilization is an artifact of top-down rulemaking rather than cumulative stance-taking. If the accusation trajectory runs parallel on platforms with and without such rules, the finding supports a grassroots-prescriptivism account over a governance-driven one. This generates two further research questions.

The first concerns tempo and mechanism: *How rapidly did the accusation stabilize, and through what measurable sociolinguistic processes?* (RQ3). Classic cases of enregisterment unfold across decades. If the same consolidation pattern appears across months in online discourse, platforms may represent a new compression factor. The second concerns the relationship between grassroots policing and formal governance: *Does the accusation stabilize independently of formal platform governance rules?* (RQ4).

Moreover, epistemic injustice is harm when a speaker is wronged in their capacity as a knower (Fricker 2007). Recent work has extended the framework to AI contexts with the model positioned as a perpetrator of injustice (Kay et al. 2024) and AI systems theorized as artificial epistemic authorities (Hauswald 2025). However, humans can wrongly attribute AI authorship to other humans. AI-mediated communication forces a choice between gullibility and blanket distrust (Sahebi & Formosa 2025). Boundary work can help us understand these issues, whereby professions compete for jurisdictional control over domains of practice (Gieryn 1983, 1999; Abbott 1988). The frame has been extended to AI shaming, read as class-marking speech that maintains the boundary between credentialed knowledge workers and those who appear to lack the standing to write fluent prose without machine help (Sarkar 2025). A broader sociology-of-cultural-production tradition describes how communities of text-based labor defend the legitimacy of their craft under technological pressure (Hesmondhalgh 2006).

Therefore, signaling theory predicts substitution and expects that harder to fake alternatives will emerge so that screening accuracy becomes the condition of producer survival. Enregisterment theory predicts stabilization, and social-epistemic and boundary-work accounts predict that the consumer response will perform social functions but do not explain why it stabilizes specifically within a time period and at scale. A final question ties these lenses together: *What social functions does the accusation perform independently of its detection accuracy?* (RQ5). Epistemic-injustice scholarship points to the harm the accusation inflicts on falsely accused writers, and boundary studies point to how status and jurisdictional claims enable accusers. RQ5 asks which of these social functions best explains the findings in real life forums.

**3. Data and methods**

Our primary data set comprises all public comments from Hacker News (HN) and 18 sampled subreddits between 1 January 2023 and 15 May 2026. Total scanned volume across the panel is c. 25 million comments, 12 million from HN and 13 million from Reddit. Reddit data was pulled from the Arctic Shift archive via the public JSON API. Hacker News data was pulled from the Algolia Hacker News search archive. Subreddit selection captures variation across four community types: AI-focused communities (r/aiwars, r/ArtistHate, r/ChatGPT, r/OpenAI, r/MachineLearning, r/LocalLLaMA, r/singularity), creative communities (r/Art, r/writing, r/books), general discourse communities (r/AskReddit, r/news, r/changemyview, r/explainlikeimfive, r/AskHistorians, r/science), and tech and academic communities (r/programming, r/AskAcademia). The sampling fraction is held constant across months for these subs, which preserves within-sub time-series comparability.

Candidate comments were identified using a 137-pattern regex lexicon organized into five tiers. Tier 1 ("Direct") captures direct accusations such as "*ChatGPT wrote this*," "*is this AI-generated*," and "*OP is a bot*." Tier 2 ("Pejorative") captures pejorative labels: *AI slop*, *GPT garbage*, *ML drivel*, *robo-writing*, and related cognates, with bare "*slop*" requiring an AI-context check. Tier 3 ("Style") captures stylistic-tell callouts including em-dash mentions, the "*delve*" callout, tricolon mentions, and the broader vocabulary of "*classic AI signature*." Tier 4 ("Mocking") captures mock and parody patterns matching canonical AI-assistant phrases ("*fellow humans*," "*in the rapidly evolving landscape*," "*rich tapestry*"). Tier 5 ("Indirect") captures indirect sense-based identification ("*smells like AI*," "*reads like ChatGPT*," "*uncanny valley of writing*"). High-false-positive patterns in Tiers 3 and 4, such as "*worth noting*," "*it's important to note*," and "*is this a human*," require an AI-context token within 250 characters of the match. The lexicon and full audit notes are available in the supplementary materials (Online Resource 1 and Online Resource 2).

To correct for the rule classifier's tier-level imprecision, two stratified per-comment LLM judgment passes were run with Claude Opus 4.7. The Reddit pass drew a 5,000-comment sample with 1,000 candidates per tier, balanced across year and sub-cluster strata. The Hacker News pass drew a separate 2,500-comment sample with 500 candidates per tier, balanced across the 41 months. Each comment in both samples was classified by per-comment LLM judgment into one of five categories. REAL covers genuine AI accusations including pejorative framing, direct callouts, tell-callouts, and accusatory questions. DISCLOSURE covers comments whose text is itself AI-generated or self-identifies as AI. NEUTRAL-REF covers non-accusatory references to AI, including on-topic discussion in AI subs and user-reported AI use. FP covers regex false positives where the pattern fired on ordinary text. AMBIGUOUS covers cases that cannot be decided from the comment alone.

Affective hardening was tested through three independent procedures. First, the proportion of Tier 2 hits whose body text contained the slop frame, the older derogatory frame set (*drivel*, *garbage*, *trash*, *vomit*, *sludge*, *mush*, *gunk*, *junk*, *crap*, *word salad*, *nonsense*), both, or neither was computed by month across all 736 Reddit sub-months. Second, the VADER (Valence Aware Dictionary and sEntiment Reasoner) sentiment compound score was computed for every LLM-validated REAL accusation on Reddit, aggregated to month and to tier. Third, a 300-thread stratified sample of LLM-validated REAL Reddit accusations was qualitatively coded into one of five speech-act types: SNEER (mocking dismissal), DISMISS (curt rejection), MOCKERY (parodic imitation), GATEKEEP (community-membership claim or rule-enforcement), and STRUCTURAL_PROTEST (objection to AI use as a general phenomenon rather than to the specific comment).

A 14-pattern placebo lexicon of pre-2022 inauthenticity vocabulary (*shill*, *astroturf*, *sockpuppet*, *paid shill*, *fake account*, *corporate shill*, *talking points*, *payola*, and variants) was applied to the same corpus on both platforms. The placebo captures inauthenticity-accusation speech that predates ChatGPT and is conceptually adjacent to AI-accusation speech but lexically distinct. If the rising trajectory of T2 hits reflected a general escalation of suspicious reading, the placebo should rise in parallel. Per-tier hits were aggregated to monthly totals. T2 share is computed as T2 hits divided by total candidate-comment count within the relevant cell. LLM-validated REAL share is computed per year from the stratified samples and applied to candidate aggregates to estimate confirmed-accusation volumes.

Finally, a two-part feature analysis was run on body prose to test the screening-accuracy prediction. Part 1 computed article density, contraction rate, formal-register adverb frequency, preposition density, sentence-length variance, and mean token length on all 4,704 labeled-sample comments with usable bodies, and compared DISCLOSURE comments to REAL comments via Mann-Whitney U tests. Part 2 pulled the parent comments for 800 sampled LLM-validated REAL Reddit accusations via the Arctic Shift comments-by-id endpoint, retained the 421 cases whose parent was itself a comment rather than a top-level post, and built a matched control sample of 2,048 non-accused comments drawn from the same sub-month with body length matched to within +/- 30 percent of each accused parent. The same six proxy markers were computed on accused parents and controls. A logistic regression of P(accused) on standardized markers, log body length, and 17 sub fixed effects tested whether the marker distribution that distinguishes AI-coded text from human accusation text also predicts accusation in the matched human-comment sample. All statistical procedures were implemented in numpy and validated against scipy/statsmodels reference values on small test cases.

**4. Findings**

RQ1 asks "*Did online readers coordinate on a substitute signal for human authorship after the diffusion of generative LLMs, and if so, what properties characterize it?*" The answer is yes. A pejorative accusation register emerged on both platforms over the study window, rose on parallel trajectories, and displaced pre-existing inauthenticity vocabulary rather than supplementing it. The headline trajectory is consistent across platforms. On Hacker News, the share of candidate comments carrying a Pejorative (T2) pattern rose from 2.5 percent in January 2023 to 26.6 percent in April 2026, with the twelve-month moving average crossing 10 percent in late 2024 and 20 percent in mid-2025. On Reddit, aggregated across all 18 subreddits, the same share rose from 1.5 percent in January 2023 to 24.4 percent in January 2026, with the rate stabilizing at 24 percent through the first four months of 2026. The two trajectories run roughly parallel, with the Hacker News share holding consistently 2 to 4 percentage points above the Reddit aggregate at each inflection point. Raw candidate volumes rose alongside the share. Reddit per-month candidate counts rose from 4,961 in January 2023 to 17,591 in February 2026, a 3.5x increase. Combined with the order-of-magnitude rise in T2 share, the absolute volume of Pejorative hits on Reddit grew from 72 per month in January 2023 to 4,290 per month in February 2026, a 60x increase. On Hacker News the equivalent absolute rise was approximately 33x, from 39 T2 hits to 1,306. Fig. 1 plots the two trajectories together with the three inflection points marked.

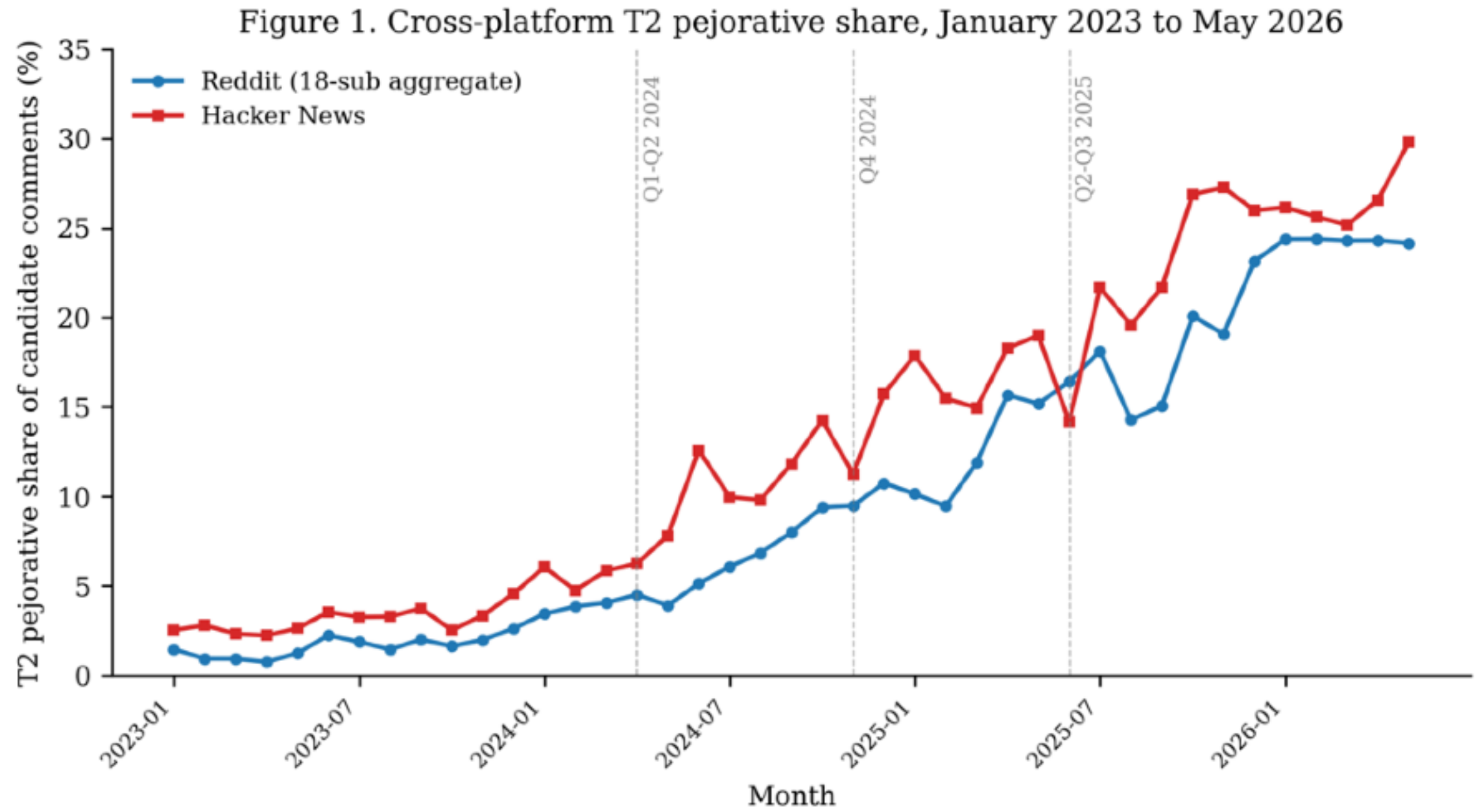

*Fig. 1 Cross-platform Pejorative (Tier 2) share, January 2023 to May 2026.*

Table 1 reports the share at eight evenly spaced months alongside the Reddit placebo share for direct visual comparison. Mann-Kendall trend tests on the full 41-month series confirm the directionality formally. Reddit T2 share rises with tau = 0.90 ($z = 8.28$, $p < 1e\text{-}15$) and Hacker News T2 share rises with tau = 0.87 ($z = 7.99$, $p < 1e\text{-}15$). The Reddit placebo share declines weakly with tau = -0.25 ($z = -2.26$, $p = 0.024$) and the Hacker News placebo share declines more strongly with tau = -0.64 ($z = -5.85$, $p < 1e\text{-}8$). A 2x2 chi-square comparing T2 versus placebo composition in 2023 against 2026 on Reddit yields chi-square = 11,116 on 1 degree of freedom (p effectively zero).

***Table 1.*** *Cross-platform Tier 2 Pejorative share by selected month, with Reddit placebo share for falsification.*

| Month | Reddit T2 share | Reddit placebo share | HN T2 share |
|---|---|---|---|
| 2023-01 | 1.5% | 8.3% | 2.5% |
| 2023-07 | 1.8% | 12.4% | 3.6% |
| 2024-01 | 3.4% | 12.2% | 6.0% |
| 2024-07 | 4.3% | 11.0% | 7.9% |
| 2025-01 | 10.1% | 15.7% | 17.9% |
| 2025-07 | 16.4% | 7.0% | 19.0% |
| 2026-01 | 24.4% | 6.9% | 26.2% |
| 2026-04 | 24.3% | 5.3% | 26.6% |

The placebo lexicon traces a flat or weakly declining share trajectory across the same window. On Hacker News, placebo counts remained in a band of 60 to 350 hits per month across all 41 months with no upward trend. On Reddit, raw placebo counts grew modestly with overall comment volume, but their share of accusation-like comments fell from 8.3 percent in January 2023 to 5.3 percent in April 2026. The pre-2022 inauthenticity register did not track the rise of the AI register. The falsification design holds at the cross-platform level, which rules out the alternative interpretation that forum readers simply became more suspicious in general over the period.

Three inflection points emerge cleanly from the time series and align across platforms. The first runs from March through June 2024, when T2 share moves from approximately 3 percent to 5 percent on Reddit and from 4 to 8 percent on Hacker News, corresponding to the public emergence of AI slop as a coherent frame in the English-language press during the second quarter of 2024. The second runs from August through December 2024, when T2 share doubles to the 8 to 11 percent range on Reddit and the 11 to 16 percent range on Hacker News. The third runs from March through July 2025, when share crosses 15 percent on Reddit and 20 percent on Hacker News.

A linear regression of T2 share on month index with sub fixed effects on the 18-sub Reddit panel yields a coefficient of 0.00663 per month (SE = 0.00018, $p < 1e-15$) and $R^2 = 0.72$. Roughly two-thirds of a percentage point per month, sustained across 41 months, is the within-panel rate at which the new register expanded. The substitute is not monolithic. The five tiers of policing speech operate simultaneously and a validation pass against LLM judgments establishes a stable internal hierarchy.

The per-tier validation against the LLM-judged 5,000-comment Reddit sample produces a stable hierarchy of confirmed-accusation rates. Tier 2 pejorative labels dominate at 78.0 percent REAL, with an additional 0.5 percent DISCLOSURE for a combined positive rate of 78.5 percent. Tier 1 direct accusations sit at 35.1 percent REAL and 38.9 percent combined, less than half the T2 rate. Tier 3 stylistic-tell callouts sit at 34.9 percent REAL and 37.8 percent combined. Tier 5 sense-based identifications sit at 28.5 percent REAL and 33.2 percent combined. Tier 4 parody patterns sit at 18.6 percent REAL, with an unusual 11.8 percent DISCLOSURE share for a combined 30.4 percent. T4 patterns sometimes catch comments that are themselves AI-generated parody or AI-assistant boilerplate, which is the only Reddit tier where DISCLOSURE share approaches double digits.

The independent 2,500-comment Hacker News LLM judgment pass replicates the hierarchy. T2 pejorative labels sit at 72.6 percent REAL, T1 direct accusations at 32.8 percent, T3 stylistic-tell callouts at 16.6 percent, T5 sense-based identifications at 14.8 percent, and T4 parody patterns at 4.8 percent. The ordering T2 > T1 > T3 > T5 > T4 holds exactly on both platforms. Absolute precision rates are lower in the lower tiers on Hacker News because the HN corpus contains a higher density of formal expository English in tech threads, which the T3 and T4 high-FP patterns fire on more often. Tier 4 in particular sits at 57 percent FP on Hacker News compared to 27 percent FP on Reddit, driven by the "fellow humans" pattern firing on sincere humanist rhetoric in political and ethical HN threads. The two platforms agree that pejorative labels are the most reliable form of AI-policing speech and that mock-parody patterns are the least. Fig. 2 plots the per-tier REAL rates side by side with Wilson 95% confidence intervals; Table 2 reports the underlying counts.

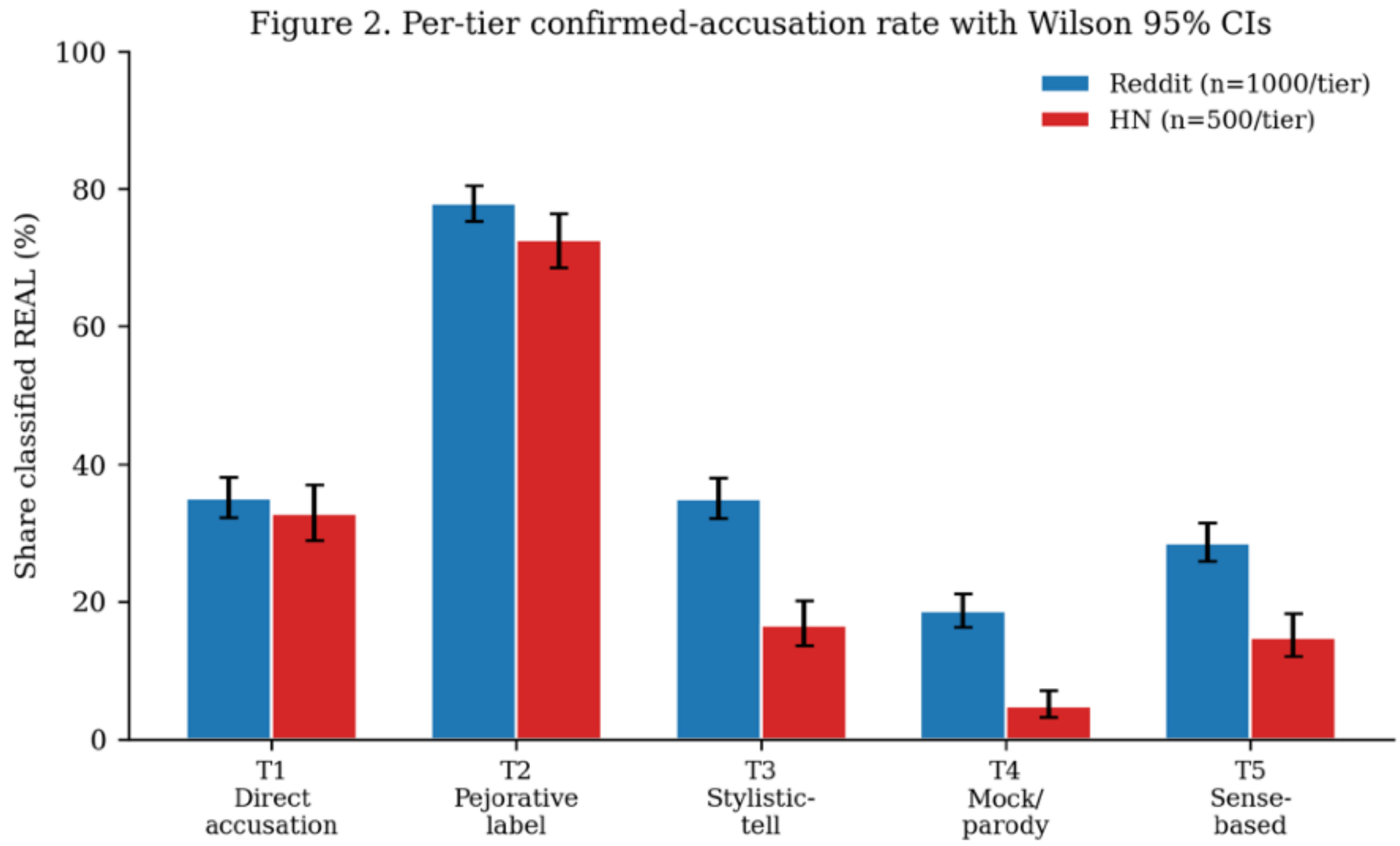

*Fig. 2 Per-tier REAL share on Reddit (n=1,000 per tier) and Hacker News (n=500 per tier) with Wilson 95% confidence intervals.*

***Table 2.*** *Per-tier precision with Wilson 95% confidence intervals, both platforms.*

| **Tier** | **Reddit n** | **Reddit REAL% [95% CI]** | **HN n** | **HN REAL% [95% CI]** |
|---|---|---|---|---|
| T1 direct accusation | 1000 | 35.1% [32.2-38.1] | 500 | 32.8% [28.8-37.0] |
| T2 pejorative label | 1000 | 78.0% [75.3-80.5] | 500 | 72.6% [68.5-76.3] |
| T3 stylistic-tell callout | 1000 | 34.9% [32.0-37.9] | 500 | 16.6% [13.6-20.1] |
| T4 mock/parody | 1000 | 18.6% [16.3-21.1] | 500 | 4.8% [3.2-7.0] |
| T5 sense-based | 1000 | 28.5% [25.8-31.4] | 500 | 14.8% [12.0-18.2] |

A logistic regression of P(REAL) on tier, year, and sub-cluster on the 5,000-comment Reddit sample confirms the picture at the population level. Tier 2 is 5.4 times more likely to be a REAL accusation than Tier 1 holding year and cluster constant (OR = 5.40, z = 16.2, $p < 1e-15$). Tier 4 is half as likely as Tier 1 (OR = 0.49, z = -6.6, $p < 1e-10$). Each one-year shift toward 2026 multiplies the odds of REAL by 1.39 (z = 9.1, $p < 1e-15$) after controlling for tier and cluster. The tech-academic cluster carries odds 2.2 times higher than the AI-focused cluster (z = 4.5, $p < 1e-5$). The parallel HN model (without cluster) returns OR = 5.94 for T2 versus T1 and OR = 0.10 for T4 versus T1, with a per-year odds multiplier of 1.66. Table 3 reports the coefficients side by side.

***Table 3.*** *Logistic regression of P(label = REAL) on tier, year, and sub-cluster.*

| **Variable** | **Reddit (n=5,000) beta** | **Reddit OR** | **Reddit p** | **HN (n=2,500) beta** | **HN OR** | **HN p** |
|---|---|---|---|---|---|---|
| Intercept | -1.17 | 0.31 | <0.001 | -1.39 | 0.25 | <0.001 |
| T2 vs T1 | +1.69 | 5.40 | <0.001 | +1.78 | 5.94 | <0.001 |
| T3 vs T1 | -0.09 | 0.91 | 0.35 | -0.95 | 0.39 | <0.001 |
| T4 vs T1 | -0.71 | 0.49 | <0.001 | -2.35 | 0.10 | <0.001 |
| T5 vs T1 | -0.30 | 0.74 | 0.002 | -1.09 | 0.34 | <0.001 |
| Year shift (per year) | +0.33 | 1.39 | <0.001 | +0.51 | 1.66 | <0.001 |
| Creative vs AI cluster | +0.41 | 1.51 | 0.006 | (not estimated) | --- | --- |
| Discourse vs AI cluster | -0.08 | 0.92 | 0.44 | (not estimated) | --- | --- |
| Tech-academic vs AI cluster | +0.80 | 2.23 | <0.001 | (not estimated) | --- | --- |
| Pseudo-R^2 | 0.150 | --- | --- | 0.265 | --- | --- |

The hierarchy explains how the register is organized. When a person uses a pejorative label (slop, drivel, garbage), the probability that the comment is a genuine accusation against suspected AI writing is 73 to 78 percent depending on platform. When the same user reaches for a stylistic-tell callout (em-dash, delve, classic GPT), the rate is closer to 17 to 35 percent. When they reach for a direct accusation, the rate is also around 33 to 35 percent because the regex captures a substantial volume of on-topic AI references in AI-focused subs and HN tech threads ("is this ChatGPT-generated") that read as honest curiosity rather than accusation. The pejorative register has become the most reliable form of AI-policing speech, far above the more analytical or direct alternatives. Short surface forms that compress identification, condemnation, and group membership into one speech act are precisely the kind of substitute that low-cost coordination should produce.

The hierarchy is stable across community types. AI-focused subs show 39 percent REAL across all tiers combined; creative subs show 48 percent; discourse subs show 30 percent REAL with 50 percent FP, reflecting the higher density of long-form expository English in those subs; tech and academic subs show 61 percent REAL, the highest of any cluster. The tier ordering (T2 > T1 ≈ T3 > T5 > T4) is preserved within each cluster, with T2 dominant in all four.

Per-year, the share of Reddit candidates that are REAL accusations rises monotonically from 18.9 percent in 2023 to 30.4 percent in 2024 to 45.3 percent in 2025 to 55.2 percent in the first four months of 2026. Applying the per-year REAL rate to per-year candidate volume gives a confirmed-accusation estimate of roughly 15,700 real Reddit accusations in 2023, 26,100 in 2024, 89,400 in 2025, and 40,200 in the first four months of 2026, which extrapolates to roughly 120,000 confirmed accusations annualized for 2026. On Hacker News the analogous per-year REAL rates are 19.5 percent (2023), 26.6 percent (2024), 32.4 percent (2025), and 43.4 percent (2026 Jan-May). Combined HN confirmed accusations total approximately 3,664 in 2023, 4,313 in 2024, 10,349 in 2025, and 10,310 in 2026 through 15 May.

RQ2 asks, *Does the substitute signal satisfy the screening-accuracy condition that signaling theory predicts is necessary for a counter-signal to survive?* The findings measure counter-signal volume rather than counter-signal accuracy. Whether a candidate comment performs the accusation of being AI-generated is distinct from whether the target prose was in fact statistically likely to be AI-generated. Human heuristics for distinguishing AI from human writing are systematically miscalibrated (Jakesch et al. 2023), and the same heuristics drive the policing speech counted here. From a signaling-theoretic standpoint the relevant quantity for RQ1 was counter-signal volume. RQ2 asks whether the substitute also does the screening work that signal theory expects. The matched-control test addresses this directly.

We computed article density, contraction rate, formal-register adverb frequency, preposition density, sentence-length variance, and mean token length on the body text of all 4,704 labeled comments with usable bodies. The means for the four LLM label classes show a clean separation between DISCLOSURE (AI-generated text or self-identified AI use) and REAL (human-authored accusations). DISCLOSURE comments carry 30 percent lower contraction rate (1.53 versus 2.18 per 100 words), 2.3 times higher formal-register adverb density (0.042 versus 0.018), 22 percent higher preposition density (9.31 versus 7.62), and 2.9 times higher sentence-length variance (299 versus 104). Mann-Whitney U tests confirm that DISCLOSURE differs from REAL on formal register ($p < 1e-9$), preposition density ($p < 1e-11$), sentence variance ($p < 1e-15$), and mean token length ($p < 1e-15$). The proxy markers distinguish AI-coded text from human accusation text at strong statistical significance. If the new accusation were tracking these features, accused parent comments should look more like the DISCLOSURE class than non-accused controls.

The second part of the test asks that question. We pulled the parent comments for 800 sampled REAL Reddit accusations, kept the 421 cases where the parent was itself a comment, and built a matched control sample of 2,048 non-accused comments drawn from the same sub-month and length range. A logistic regression of P(accused) on standardized markers, log body length, and 17 sub fixed effects gives the result in Table 4. None of the four markers that distinguished DISCLOSURE from REAL in the descriptive pass significantly predicts accusation in the matched-control test. Formal-register adverb density carries a positive but only marginally significant coefficient ($p = 0.083$).

***Table 4.*** *Logistic regression P(accused) on standardized proxy markers and log length, with 17 sub fixed effects. n = 421 accused parents and 2,048 matched controls. OR is the odds ratio per one SD increase in the marker.*

| **Variable** | **beta** | **SE** | **z** | **p** | **OR per SD** |
|---|---|---|---|---|---|
| Intercept | -1.39 | 0.50 | -2.76 | 0.006 | 0.25 |
| Article density | -0.10 | 0.06 | -1.62 | 0.105 | 0.91 |
| Contraction rate | +0.02 | 0.06 | 0.32 | 0.750 | 1.02 |
| Formal register | +0.07 | 0.04 | 1.74 | 0.083 | 1.08 |
| Preposition density | -0.06 | 0.06 | -1.08 | 0.281 | 0.94 |
| Sentence-length variance | +0.01 | 0.05 | 0.10 | 0.917 | 1.01 |
| Mean token length | -0.25 | 0.07 | -3.67 | <0.001 | 0.78 |
| Log body length | +0.19 | 0.06 | 3.21 | 0.001 | 1.21 |

The two-part result is that the proxy markers separate AI-generated text from human-written text, but the same markers do not separate accused human comments from matched non-accused human comments. The new accusation stabilized and performed the social work of a substitute signal without doing the screening work that the standard signaling theory account would have predicted it must to socially survive. Accusers, at the population level, make their accusations without reference to the prose features that statistically indicate AI-generated content, relying instead on specific phrases, punctuation patterns, content cues, contextual signals about who the writer is, or heuristics that are systematically miscalibrated with respect to the underlying statistical signature of AI text (Jakesch et al. 2023).

*How rapidly did the accusation stabilize, and through what measurable sociolinguistic processes?* Over the same window, the share of T2 hits using older pejoratives (drivel, garbage, trash, vomit, sludge, mush, gunk, junk, crap, word salad, nonsense) fell from 31.6 percent in 2023 to 22.7 percent in 2024 to 10.9 percent in 2025 to 8.4 percent in 2026. The slop frame did not simply rise alongside older pejoratives. Slop displaced them. By 2026 the older derogatory frame set captures fewer than one in twelve pejorative-label accusations. A pejorative-label tone that in 2023 ran on a half-dozen frames now runs almost entirely on one. The consolidation is what coordinated substitution looks like at the lexical level: a population converging on a single dominant form that minimizes coordination cost and maximizes mutual recognizability. Figure 3 plots the two trajectories together.

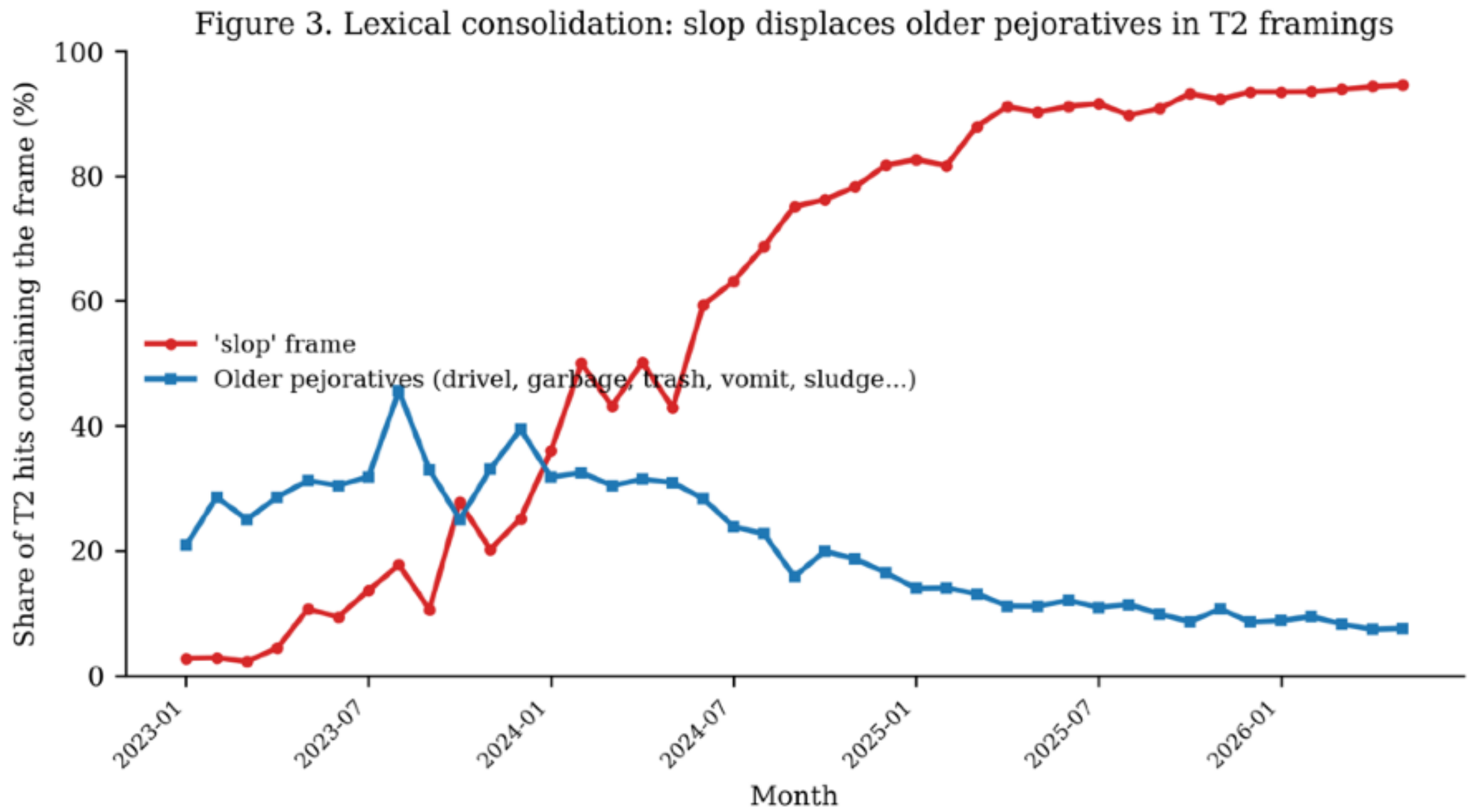

*Fig. 3 Lexical consolidation: slop displaces older pejoratives in Pejorative (Tier 2) framings.*

The second measure is sentiment. The mean VADER compound score on the 1,950 LLM-validated REAL Reddit accusations fell from +0.21 in 2023 to +0.08 in 2024 and held at +0.07 through 2026. The share of accusations with compound score below -0.3 rose from 16 percent to 28 percent over the same window, and the share with score below -0.6 rose from 7.5 percent to 17.4 percent. The mean tone of accusation speech became measurably more negative as the accusation matured, and the share of strongly negative accusations more than doubled. The sentiment shape lines up with the tier hierarchy: Direct T1 accusations carry the most negative mean compound (-0.03), T2 pejorative labels sit at +0.04, T3 stylistic-tell callouts at +0.14, T5 sense-based identifications at +0.19, and T4 parody at +0.24. Parody is the least negative because it imitates rather than condemns. An OLS regression of compound on tier and year shows that all four tier dummies are significantly more positive than the T1 reference (T2 +0.09, $p = 0.02$; T3 +0.18, $p <$ 1e-4; T4 +0.27, $p <$ 1e-6; T5 +0.23, $p <$ 1e-6). The continuous year term is small and not significant in the joint model (beta = -0.021 per year, $p = 0.18$), which means the apparent year-on-year drop in mean sentiment is largely mediated by the shifting tier mix rather than by within-tier hardening. As the accusation matures, more accusations are T2 (mid-affect) and fewer are T1 (highest-affect), and the population mean drops accordingly.

Within Reddit the cluster patterns add to the picture. The three clusters that scaled fastest (AI-focused, creative, tech-academic) reached their inflection points three to six months ahead of the discourse cluster. r/aiwars rose from 4 percent T2 share in January 2023 to 29 percent in April 2026, tracing the same three-inflection-point shape as the overall aggregate. r/ArtistHate, founded December 2022 and effectively absent from the January 2023 panel, reached 16 percent T2 share by January 2024 and stabilized at 37 to 45 percent through 2025-2026. r/writing rose from 2 percent to 44 percent T2 share, the steepest single-sub shift in the panel. r/programming rose from 1 percent to a 2025 peak of 42 percent before settling at 31 percent in April 2026, the highest sustained tech-cluster signal. The discourse cluster lagged by roughly twelve to eighteen months. r/AskHistorians went from 0 percent T2 share in January 2023 to 26 percent in April 2026, with the inflection happening in mid-2025. r/AskAcademia, r/news, and r/AskReddit followed similar patterns at lower amplitude.

A third stabilization measure, the migration of speech-act types over the study window, is reported under RQ5 below because its primary analytic contribution concerns the social functions the accusation performs rather than the pace of stabilization per se. The temporal pattern (mockery declining from 26 percent to 7 percent of accusations while structural protest tripled) is consistent with the lexical and sentiment evidence of a register that consolidated and hardened over 41 months.

GATEKEEP rose from a near-absent 1.9 percent in 2023 to 16.5 percent in 2026, driven in part by formal moderator-removal notices on r/changemyview, r/programming, and r/AskHistorians, consistent with community-enforcement institutionalization. SNEER did not show a clean linear rise: 18.5 percent in 2023, 14.0 percent in 2024, 18.3 percent in 2025, 15.3 percent in 2026. The absolute count of SNEER doubled and the tone of those jeers intensified, but its share was held down by the larger growth in STRUCTURAL_PROTEST. The composite is one of register migration. Dismissive one-liners and parody gave way to broader political grievances about AI and to institutionalized community-rule enforcement, with sneers remaining as a steady undercurrent rather than the dominant mode. Figure 4 plots the year-by-year distribution as stacked shares; Table 5 reports the three hardening legs together.

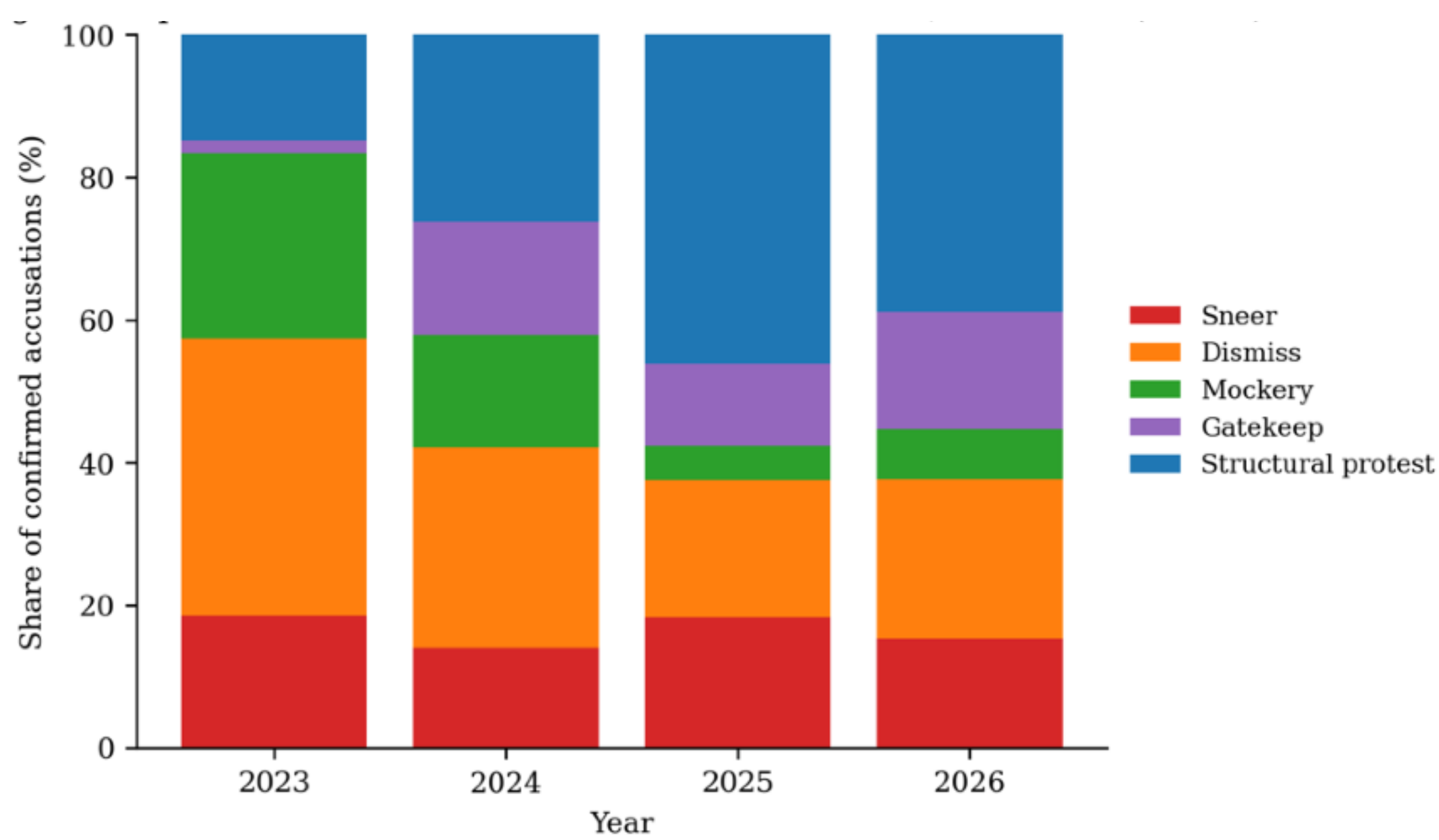

*Fig. 4 Speech-act distribution of confirmed accusations, 2023-2026, n=300.*

***Table 5.*** *Three independent measures of affective hardening, by year.*

| Year | Slop share of T2 | Mean VADER compound | MOCKERY share | GATEKEEP share | STRUCTURAL_PROTEST share |
|---|---|---|---|---|---|
| 2023 | 12.9% | +0.21 | 25.9% | 1.9% | 14.8% |
| 2024 | 66.7% | +0.08 | 15.8% | 15.8% | 26.3% |
| 2025 | 90.7% | +0.07 | 4.8% | 11.5% | 46.2% |

| 2026 | 93.8% | +0.07 | 7.1% | 16.5% | 38.8% |
|---|---|---|---|---|---|

*Does the accusation stabilize independently of formal platform governance rules?* Comment-level policing operates whether or not platforms enforce explicit rules. The number of subreddits with explicit AI rules doubled between July 2023 and November 2024 (Lloyd et al. 2025). Our comment-level data shows what that change meant for accusations of AI use. 17 of 18 subs reached double-digit T2 share by 2026, including subs with formal AI rules (r/aiwars, r/ArtistHate, r/Art, r/writing, r/books) and subs without them (r/AskReddit, r/news, r/programming). A subset of the panel is composed of subs whose entire purpose is the rejection of AI: r/ArtistHate (founded December 2022) and to a lesser extent r/aiwars. By 2025-2026 these collectively produced roughly 9 percent of all candidate hits on Reddit despite representing well under 1 percent of underlying comment volume. AI-policing has migrated from individual posts into community-level institutions, with the same lexicon operating at both scales.

r/changemyview is an outlier, holding below 6 percent T2 share throughout the panel. The most likely explanation is the subreddit's explicit rule structure: a long-standing community norm that incentivizes good-faith engagement, a separate AI-disclosure rule, and active moderation. Where platform structure already enforces a norm of substantive reply, the new AI accusation register has less work to do, and stigmatizing speech finds fewer targets. On Hacker News, which lacks subreddit-style rule infrastructure entirely, the trajectory runs 2 to 4 percentage points above the Reddit aggregate at every inflection point. The cross-platform parallel supports a grassroots-prescriptivism account over a governance-driven one.

*What social functions does the accusation register perform independently of its detection accuracy?* The matched-control null result establishes that the accusation register does not track the prose features that distinguish AI text from human text. Something other than screening accuracy sustains it. Three bodies of evidence bear on what that something is: the speech-act distribution and its temporal migration, the cluster-level variation in accusation ecology, and the trajectory of the DISCLOSURE category.

A 300-thread stratified sample of LLM-validated REAL Reddit accusations was coded into five speech-act categories. The aggregate distribution runs STRUCTURAL_PROTEST (34.7 percent), DISMISS (25.3 percent), SNEER (16.7 percent), GATEKEEP (12.0 percent), and MOCKERY (11.3 percent). There was a substantial shift in category use over time. MOCKERY accounted for 25.9 percent of 2023 accusations and fell to 4.8 percent by 2025 and 7.1 percent by 2026, consistent with parody becoming less central as the accusation hardened. STRUCTURAL_PROTEST tripled from 14.8 percent in 2023 to 46.2 percent in 2025 before settling at 38.8 percent in 2026, reflecting a migration of the accusation act from individual-comment policing toward broader phenomenon-level objections.

The early dominance of mockery and dismissal is consistent with individual-level boundary policing: writers marking other writers as outsiders through parody and contempt. The rise of structural protest reflects a shift from policing individual comments to contesting the broader phenomenon of AI-generated content in public discourse, a move from micro-level stance-taking to macro-level political speech. The rise of gatekeeping reflects the institutionalization of that policing into community norms, with moderator-removal notices and rule-based enforcement replacing ad hoc sneering. As the accusation's social work matured, the functions it performed shifted from status marking and boundary defense toward collective political action and institutional norm enforcement.

The DISCLOSURE share among Reddit candidates fell from 11.5 percent in 2023 to 2.8 percent in 2026. The HN DISCLOSURE share fell from 5.6 percent to 4.5 percent over the same window. Two readings are plausible, both requiring future study. The first is that early AI output was easier to identify as AI output, both because models produced more characteristic boilerplate and because users tended to paste rather than rewrite. The second is that AI-style writers learned to disguise their prose or moved to communities with weaker enforcement norms as forum-level policing intensified. Moreover, detection-improving technology is unlikely to fix accusation accuracy, because the accusation is not anchored to the detection problem in the first place; its function is social gatekeeping.

**5. Discussion**

Our findings show that, when generative AI degraded the signal value of good prose as a marker of effort, populations of online forum readers coordinated on a substitute accusation register. These signals stabilized through enregisterment but did not acquire the detection accuracy that classical signaling theory would have predicted as the condition for the substitute's survival. The substitute survived instead on other grounds, namely that it was generally effective as a gatekeeper for any speech that a commenter wanted silenced (for whatever reason). The placebo design rules out generalized suspicion, while the three measures of affective hardening rule out the possibility that the register is decomposing, and the cross-platform parallel structure rules out platform-specific drift. To our knowledge, this is the first observational empirical finding that AI-use accusations are not made primarily to identify the use of AI.

The persistent 2-4 percentage point lead Hacker News carries over Reddit across all 41 months is not incidental. HN's user base is disproportionately composed of software engineers, researchers, and early adopters who encountered LLMs before the general public. These users thus had both greater prior exposure to AI-generated text and stronger prior motivation to identify and police it. The gap is consistent with a diffusion model in which technically fluent communities develop and stabilize accusation norms earlier, with those norms spreading to broader-audience platforms subsequently. Reddit's convergence on nearly the same share by 2026 confirms that the register is not platform-specific and the HN lead likely reflects timing.

The implication for signaling theory concerns the receiving side of an unpriced discourse. The sending side of the signal collapsed under generative AI, a phenomenon that has been modeled in a priced labor market (Galdin & Silbert 2025). Under conditions where the underlying detection problem cannot be solved at the lay level, substitute signals can stabilize on social purpose alone without acquiring technical accuracy. The purpose comes from the social functions the accusation performs, each of which gives selection pressure that does not depend on whether the accusations track the actual statistical likelihood for AI use. The conditions under which such equilibria can be expected to form are common in domains touched by generative AI. The empirical map provided here is a starting point for theorizing them. Our results here would predict that similar AI-use accusations will form for image authentication, voice authentication, and code authorship among others, with the core intent of the lay accusation being gatekeeping rather than empirically accurate detection of AI use. This will become increasingly problematic as AI in those areas reduces even the empirically detectable tip-offs that experts can find. This could have the effect of increasing the role of experts in verifying AI vs. non-AI content; or it could greatly reduce trust in any type of medium that can be plausibly generated by AI.

Our findings also invert the potential direction of harm through the recent literature on epistemic injustice and AI. Existing extensions of Fricker's framework position the AI system as a perpetrator of testimonial injustice, with harm flowing from machine to human (Kay et al. 2024), while adjacent work theorizes that AI systems are artificial epistemic authorities (Hauswald 2025). Our findings flow in the opposite direction, suggesting that lay AI literacy, operating in degraded epistemic conditions, produces testimonial injustice between humans at population scale, in this case directed towards writers by readers. The structural dilemma has been articulated philosophically (Sahebi & Formosa 2025), and the present study supplies the empirical mechanism by which blanket distrust gets operationalized as a lay accusation. The Fricker tradition's standard remedies of epistemic virtues at the level of the individual hearer and hermeneutical resources at the level of the community need supplementing when the producer of testimonial injustice is coordinated user activity of a stabilized accusation rather than an institution or an interlocutor.

Testimonial injustice in Fricker's original account requires that a speaker's credibility be deflated because of their perceived identity like their race, gender, or social position as the hearer construes it (Fricker 2007). What the accusation data document here is credibility deflation based on perceived authorship method, not identity as such. This may sit closer to hermeneutical injustice: accusers lack the interpretive resources to accurately assess whether a text is AI-generated, and that gap produces harm not by intent but by structural incapacity. It may also constitute a novel category, one in which the hermeneutical deficit is technical rather than social: the community does not lack vocabulary for the experience; it faces a detection task that is, at the lay level, currently unsolvable. These two framings suggest different remedies, and future work should establish which fits the empirical mechanism more precisely before recommending a course of corrective action.

The third contribution of this research speaks to the sociology of cultural production. AI shaming has been read conceptually as boundary work maintaining class solidarity among credentialed knowledge workers, drawing on Gieryn for the boundary-work mechanism (Sarkar 2025; Gieryn 1983, 1999). If the accusations were tracking AI signals, they would predict accusation in the matched sample. They do not, and the temporal pattern of the speech-act coding (mockery collapsing, gatekeeping rising, structural protest tripling) gives a picture of what kind of social work the accusation does as it matures. Communities of text-based labor (developers, writers, academics, hobbyist forum participants) may be coordinating on accusations that assert the value of human authorship, but those accusations are not currently doing the screening work that would ground that assertion.

The rise of gatekeeping that the data indicates sits at the boundary between grassroots and institutional policing. Gatekeeping rose from a near-absent 1.9 percent in 2023 to 16.5 percent of confirmed accusations in 2026. A substantial share of late-period gatekeeping speech consists of standardized moderator-removal notices on r/changemyview, r/programming, and r/AskHistorians, with templated language about AI rules and community norms. These comments are technically posted by moderators acting under formal rule structures, but they enter the comment stream at the same surface as ordinary user accusations and are read as part of the same accusation. The doubling of explicit AI rules on Reddit between mid-2023 and late 2024 has been documented elsewhere (Lloyd et al. 2025). The institutional layer feeds back into the comment-level accusation by 2026, with the same vocabulary operating at both scales. Grassroots and institutional policing are the same substitute observed at two scales, with the lexicon migrating freely between user comments and moderator notices. The "New Governors" framing reads platforms as quasi-governmental rule-makers, with user norms as inputs to corporate decision-making (Klonick 2018). The present data invert that flow for the AI-content domain. The macro-norm vocabulary is grassroots in origin and appears in both user comments and formal moderator notices.

There are a variety of practical implications of our findings as well. For platform designers, rule-based AI policies such as explicit bans, disclosure requirements, and moderator-removal frameworks cannot substitute for the underlying detection problem, but they can redirect accusation energy. The r/changemyview pattern suggests that communities with strong prior norms of substantive engagement are more resistant to accusation drift. Investing in those norms upstream is more tractable than adjudicating individual accusations downstream. For writers, the matched-control result means that being accused has no reliable connection to writing in a way that statistically resembles AI output. The writers most exposed are those whose prose is formal, polished, and low in contractions, which happens to match lay heuristics for AI text, regardless of how those features were produced and of whether they actually reflect AI use. The accusation says more about the community's epistemic state than about the text. For policymakers, the key implication is that mandating better AI detection tools will not resolve the social dynamics documented here. If the accusation's fitness derives from its social function rather than its detection accuracy, improved detection adds nothing to that fitness. The policy problem is an epistemic infrastructure problem, and it calls for solutions at that level through AI literacy investment, credentialing mechanisms that survive the cheap-prose era, and discourse space design that does not reward gatekeeping over engagement.

The current study has several limitations. The lexicon was tuned through one iterative false-positive audit on r/explainlikeimfive; other subs may carry idiosyncratic false-positive patterns we have not surfaced, partially mitigated by per-tier LLM precision. Platform coverage stops at Hacker News and Reddit. A pilot attempt at Stack Exchange was deferred because of access constraints, yet Stack Exchange remains a useful comparison given its formal-register expectations and active moderation. Cross-language coverage is absent, and this limitation matters more here than in comparable studies. The English-language accusation register documented here is lexically specific: "slop" carries connotations of organic waste and low-quality mass production that do not translate straightforwardly. Whether equivalent registers have stabilized in French, German, Mandarin, Arabic or other online communities, and whether non-English communities converge on a single dominant pejorative as English did (or remain more fragmented across competing terms) is an open empirical question. The answer would test whether the lexical consolidation dynamic is a general feature of online community behavior under generative AI pressure or an artifact of English-language platform demographics. The causal identification of writer-side conscious avoidance requires survey instruments or panel data tracking individual writers over time. An important next step would be a study that builds a writer-side panel tracking users who have been publicly accused to see whether their prose features change after the accusation. A second direction would be to test whether the signaling-substitution-without-accuracy claim generalizes to other domains where generative AI has degraded a previously trusted signal.

## 6. Conclusion

The post-ChatGPT period produced a test of what populations do when a signal for quality collapses and the underlying detection problem cannot be solved at the lay level. Between 2023 and 2026, readers on Hacker News and Reddit coordinated on a substitute screening signal. The substitute stabilized without acquiring detection accuracy, and its social functions supplied the fitness that classical signaling theory would have expected to come from improved screening. The matched-control test makes the miscalibration explicit, and the placebo, cross-platform, and affective-hardening checks rule out the principal alternative explanations.

The findings have implications across three literatures. For signaling theory, they describe a class of equilibria in markets with information asymmetry where substitute signals stabilize on social fitness without acquiring screening accuracy. For social epistemology, they document an inverted form of testimonial injustice in which lay AI literacy produces harm to human writers. For the sociology of cultural production, they supply the empirical anchor that boundary-work readings of AI shaming have lacked. Future work needs to take the receiving side seriously as the primary site of analysis, with production effects as downstream consequences of policing intensity. The empirical map provided here shows how that cost has shifted and where it concentrates.

**Conflict of Interest Statement**
The authors declare no financial or personal conflicts of interest that could have influenced the findings reported in this article. This study received funding from the University of Oslo, where one of the co-authors is affiliated. This study used Claude Opus 4.7 (Anthropic) to perform per-comment classification of sampled accusation data. The authors have no financial relationship with Anthropic and received no compensation or support from the company. The choice of model was made on methodological grounds; results were validated against independent coding procedures as described in Section 3.

**Data Availability Statement**
The full regex lexicon, audit notes, and supplementary coding materials are available from the corresponding author upon reasonable request. The Hacker News data was retrieved from the Algolia Hacker News Search Archive, which is publicly accessible. Reddit data was retrieved from the Arctic Shift archive via the public JSON API. Due to Reddit's terms of service, raw comment text cannot be redistributed; researchers can replicate the dataset by querying the Arctic Shift archive using the comment IDs provided in the supplementary materials. All statistical procedures were implemented in numpy and validated against scipy/statsmodels; analysis code is available from the corresponding author upon reasonable request.